\documentclass[conf]{new-aiaa} %for conference proceedings
\usepackage[utf8]{inputenc}
\usepackage{graphicx}
\usepackage{blindtext}
\usepackage{subfiles}
\usepackage{float}
\usepackage{caption}
\usepackage{array}
\usepackage{rotating}
\usepackage{makecell, multirow}
\usepackage{subcaption}
\usepackage[dvipsnames,table,x11names]{xcolor}
\usepackage{amsmath}
\usepackage[version=4]{mhchem}
\usepackage{siunitx}
\usepackage{longtable,tabularx}
\title{Underactuated Two Stage CubeSat Control Law}

\author{Maxwell Cobar\footnote{mjc1521@jagmail.southalabama.edu, Graduate Research Assistant, AIAA Student Member.} and Carlos Montalvo\footnote{cmontalvo@southalabama.edu, Associate Professor, AIAA Senior Member.}}
\affil{William B. Burnsed Jr. Department of Mechanical, Aerospace and Biomedical Engineering\\ College of Engineering\\University of South Alabama \\150 Student Services Dr. Mobile, AL, 36688}

\begin{document}
\maketitle

\begin{abstract}
This paper details the implementation of a two stage underactuated control law for a CubeSat to reduce all angular rates to zero. This CubeSat has a diagonal inertia matrix where the principal axes of inertia are not equal. This is needed for momentum transfer between axes as shown by previous work in this area. The momentum transfer will provide control to the uncontrolled axis to reduce all angular rates of a CubeSat to zero. The underactuated control utilizes actuators on only two axes but still reduces all three angular rates to zero. The first stage utilizes one axis to control the uncontrolled axis to zero and then the second stage reduces the other two axes to zero. This control law is similar to other piece wise control laws but the difference lies in only using one axis to control the third axis. The control mechanism is an integrated two axis propulsion system. The control law is derived to show its viability and a fully non-linear six degree of freedom simulation tool is utilized to verify the derivation. The simulation tool is developed by the Facility of Aerospace Systems and Technology. Simulation results are shown for a 1U, 2U, and 6U CubeSat with three and two axis control. The two stage underactuated control law is compared to a proportional controller as well as a feedback linearization control law for comparison. 
\end{abstract}

\section*{Nomenclature}
{\renewcommand\arraystretch{1.0}
\noindent\begin{longtable*}{@{}l @{\quad=\quad} l@{}}
$p,q,r$ & Components of the Angular Velocity in the Body Frame ($\frac{rad}{s}$) \\
$\dot{\vec{\omega}}$ & Angular Acceleration Vector in the Body Frame ($\frac{rad}{s^2}$) \\
$\dot{p},\dot{q},\dot{r}$ & Components of the Angular Acceleration in the Body Frame ($\frac{rad}{s^2}$) \\
$\vec{M}$ & Total Moment Vector in the Body Frame ($N \cdot m$) \\
$L,M,N$ & Components of the Total Moment in the Body Frame ($N \cdot m$) \\
$l,w,d$ & Length, Width, and Depth ($m$) \\
$m$ & Mass ($kg$) \\
$\textbf{I}$ & Mass Moments of Inertia Matrix ($kg \cdot m^2$) \\
$I_x,I_y,I_z$ & Components of the Mass Moments of Inertia ($kg \cdot m^2$) \\
${\bf S}(\vec{r})$ & Skew Symmetric Matrix Operator \\
$ADCS$  & Attitude Determination and Control Subsystem \\
$DOF$  & Degrees of Freedom \\
$GNC$  & Guidance, Navigation, and Control \\
$HEO$  & High Earth Orbit \\
$LEO$  & Low Earth Orbit \\
$MarCO$  & Mars Cube One \\
$RCS$ & Reaction Control System \\
$RW$ & Reaction Wheel \\
$ECI$  & Earth Centered Inertial Frame
\end{longtable*}}

\section{Introduction} \label{Section:Intro}
%Start with 3-axis control
%Rank deficiency LEO w/ magTs
%Problem w/ deep space - no magTs
%Other papers w/ 2-axis control
%2-axis 2-stage FL control law

Created in 1991, CubeSats are satellites with a standardized size and form factor that typically fall under the nanosatellite (NanoSat) classification. The standard size of a CubeSat is a 1 unit (1U) cube that measures 10x10x10 centimeters (cm). These cubes can be stacked to create a larger CubeSat to complete more complex missions. Due to their small size, CubeSats are usually launched into a low Earth orbit (LEO) to perform scientific research and space technology demonstrations for larger satellite projects \cite{NASA_CubeSat}. However, through miniaturization of technologies and reduced cost, CubeSats have been used for deep space missions. In May of 2018, the National Aeronautics and Space Agency’s (NASA) Mars Cube One (MarCO) interplanetary nano-spacecraft mission showed the possibilities of CubeSat utilization for deep space and helped to the develop technologies to work within a CubeSat’s small volume \cite{MARCO}. To get to its destination, CubeSats “piggyback” as secondary payloads on larger missions, but once the CubeSat is ejected, it must rely on the components within its small three-dimensional footprint \cite{DeepSpace_CubeSats}. Due to its limited volume, the components of a CubeSat must be chosen with care to ensure components will fit.

The components of a CubeSat fall into the different subsystems that make up the CubeSat system. A primary subsystem of a CubeSat is the Guidance, Navigation, and Control (GNC) subsystem. The GNC subsystem primarily deals with components used in the Attitude Determination and Control System (ADCS) \cite{NASA_GNC}. The ADCS is particularly limited by this mass and volume restriction of a CubeSat. The configuration of the ADCS is dependent on the satellite’s mission; however, it is composed of sensors, actuators, and computational hardware that can consume a large portion of the CubeSat's mass and volume budget. This is especially true when the CubeSat requires redundant actuation through a combination of reaction wheels (RWs), magnetorquers, and/or a reaction control system (RCS). A traditional control scheme for a CubeSat in low Earth orbit (LEO) is to use three magnetorquers to control the orientation and pointing direction. Using three magnetorquers allows for control over the three rotational axes of the CubeSat. However, when a CubeSat enters deep space it becomes too far away from the strong part of Earth's magnetic field to use magnetorquers. In this case, reaction wheels are used to point and control the attitude of the CubeSat. Although RWs are well suited for CubeSats due to their performance and small size, they come with a large drawback of momentum storage saturation \cite{RW_Problems}. To counteract this drawback, secondary actuation must be applied to create an external torque relative to the reaction wheel to counteract the torque created when slowing down the reaction wheel. In LEO if reaction wheels are used, desaturation can occur through the use of magnetorquers as the secondary actuation. However, in deep space or high Earth orbits (HEO) when magnetorquers are not viable, desaturation is usually accomplished by an RCS. When using a three-axis RCS with a primary propulsion system, a large demand is created on the mass, volume, and power budgets of a CubeSat \cite{ABEX_GNC}.

To prevent this large demand on the mass, volume, and power budgets, a two axis RCS system could be used. This would require an underactuated controller where the number of degrees-of-freedom (DOF) is greater than the number of actuators\cite{underactuated}. The utilization of an underactuated control law allows for an actuator failure to occur while providing a system with adequate control authority. Underactuated systems can also be used to reduce mechanical complexity and the cost of a system. For the cases described, the difficulty in controlling an underactuated system comes from the stabilization of the dynamics by any smooth or continuous time-invariant feedback law. This is due to the equations of motion violating Brockett's necessary condition, even in the case of a system that is small time locally controllable (STLC) \cite{underactuated_spacecraft_control}. Control laws for time-invariant fully actuated systems are well understood \cite{Rigid_Control}. Much work has been done to develop control laws for such complex systems, especially in terms of attitude stabilization. The first analysis of the equations of rotational motion have been addressed for rigid bodies with one, two, and three control torques in \cite{Spacecraft_Attitude_Control}.

Underactuated control for detumbling has been addressed in previous studies. These studies have explored using magnetorquers in LEO to detumble reaction wheels. The University of South Alabama has looked at using two magnetorquers to detumble the reacton wheels on a 2U CubeSat. A simple porportional control scheme reduces the performance of the detumbling maneuver but allows the CubeSat to carry one less magnetorquers \cite{Detumble_2U_CubeSat}. Kyushu University has expanded on this work by using one magnetorquer. Both of these underactuated control laws take advantage of a non-constant magnetic field when in an inclined orbit. The Kyushu University introduces a more complex control method with one magnetorquer by using Model Predictive Control and compares it with a B-dot control algorithm to show the feasibility of an underactuated control \cite{MPC_MagT}.  Horri and Hodgart proposed underactuated reaction wheel control based on Rodriguez parameterization of the attitude. They noted that the initial angular momentum must be small or otherwise accounted for by magnetorquers \cite{2RW_Control}. 

Some previous studies have discussed discontinuous feedback control methods and shown that non-symmetric inertia matrices allow for accessible attitude dynamics if the uncontrolled principal axis is not an axis of symmetry \cite{Two_Control_Torque,ActuatorFailures}. If a CubeSat has a non-uniform diagonal inertia matrix, the CubeSat can transfer momentum from the uncontrolled axis to the axis that is controlled and vice-versa. The work by \cite{Two_Control_Torque} stabilizes and controls all three axes using a series of manuevers that starts with nulling the angular rates. This is done by first nulling the two controlled axes and then using the two axes to null the 3rd axis while simultaneously ensuring that the original two axes are zero in finite time. The control law presented here commands the second axis to a constant angular rate and utilizes the 3rd axis to control the first axis. Once the first axis has reached zero, the second two axes are driven to zero. Although the method is similar it is subtly different than previous work in this area. The control law presented here is simulated on a 1U, 2U, and 6U CubeSat with a circular orbit about the equator. The main propulsion system for the CubeSat is an integrated propulsion system which provides three-axis control, but will be limited to two-axis control (roll and pitch) when showing results of the underactuated scenario. 

\section{CubeSat Dynamics} \label{Section:6DOF}
%Images of 1U,2U,6U
%Show Ix,Iy,Iz relationship
%2U on side Iy=Iz but 2U upright Iy != Iz
%Expand pdot,qdot,rdot with diagonal inertia and L,M,N != 0

\begin{figure}[H]
\centering
\includegraphics[width=0.8\textwidth]{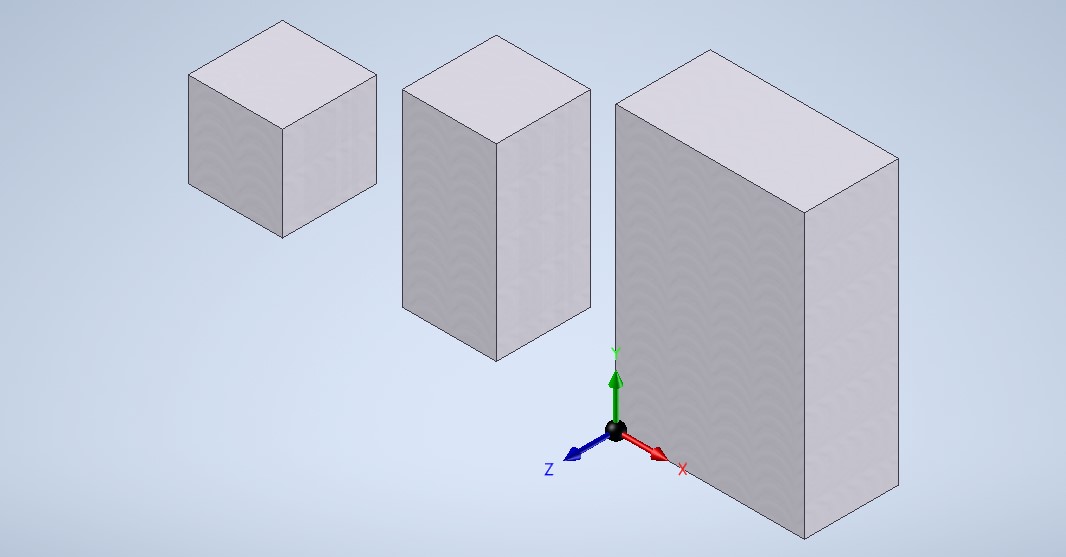}
\caption{Rudimentary CAD Model of a 1U, 2U, and 6U CubeSat}
\label{fig:CubeSats}
\end{figure}

%Before deriving the control laws, the dynamics of the CubeSat need to be explored. 
For this paper, a 1U, 2U, and 6U CubeSat with a diagonal inertia matrix are simulated. As mentioned previously, a 1U CubeSat is a 10x10x10 cm cube. The 2U CubeSat is 1U wide, 1U long, and 2U tall. A 6U CubeSat is 1U wide, 2U long, and 3U tall. A rudimentary CAD model of the different size CubeSats can be seen in Fig. \ref{fig:CubeSats}. The 1U CubeSat is on the left, the 2U is in the center, and the 6U is on the right.

\noindent In this paper, CubeSats are assumed to be solid cuboids, the mass moments of inertia are thus calculated using the following equations where $I_x,I_y,I_z$ are the mass moment of inertia along their respective axes, m is the mass, l is the length, w is the width, and d is the depth.

\begin{equation} \label{e:inertia} 
  \begin{gathered}
    I_x = \frac{m}{12}(l^2 + w^2) \\
    I_y = \frac{m}{12}(l^2 + d^2) \\
    I_z = \frac{m}{12}(d^2 + w^2)
  \end{gathered}
\end{equation}

\noindent By finding the mass moments of inertia for a CubeSat, the rotational dynamics of the CubeSat can be calculated. The rotational dynamic equation for the CubeSat is shown below where ${\bf I}_C$ is the inertia matrix of the CubeSat, $\vec{M}$ is the total moment vector, ${\bf S}(\vec{\omega})$ is the skew symmetric matrix of the angular velocity vector, and $\vec{\omega}$ is the angular velocity vector.

\begin{equation}\label{e:pqrdot}
  \dot{\vec{\omega}} = {\bf I}_C^{-1} (\vec{M} - {\bf S}(\vec{\omega}) {\bf I}_C \vec{\omega})
\end{equation}

\noindent Eqn. \ref{e:pqrdot} can be further expanded as seen below where $L,M,N$ are the components of the total moment acting on the CubeSat and p,q,r are the components of the angular velocity of the CubeSat.

\begin{equation}\label{e:pqrdot_expanded}
  \begin{Bmatrix} \dot{p} \\ \dot{q} \\ \dot{r} \end{Bmatrix} = {\bf I}_C^{-1}\left(\begin{Bmatrix} L\\M\\N\end{Bmatrix}-\begin{bmatrix} 0 & -r & q \\ r & 0 & -p \\ -q & p & 0 \end{bmatrix} {\bf I}_C\begin{Bmatrix} p\\q\\r\end{Bmatrix}\right)
\end{equation}

\noindent When the CubeSat is fully actuated, the components $L,M,N$ of the external moment vector do not equal zero. Assuming the CubeSat is experiencing angular motion about all three axes, Eqn. \ref{e:pqrdot_expanded} can be simplified as seen below.

\begin{equation} \label{e:pqrdot_diagonal} 
  \begin{gathered}
    \begin{Bmatrix} \dot{p} \\ \dot{q} \\ \dot{r} \end{Bmatrix} = \begin{bmatrix} \frac{1}{I_{xx}} & 0 & 0 \\ 0 & \frac{1}{I_{yy}} & 0 \\ 0 & 0 & \frac{1}{I_{zz}} \end{bmatrix} \left(\begin{Bmatrix} L\\M\\N\end{Bmatrix}-\begin{bmatrix} 0 & -r & q \\ r & 0 & -p \\ -q & p & 0 \end{bmatrix} \begin{bmatrix} I_{xx} & 0& 0 \\ 0 & I_{yy} & 0 \\ 0 & 0 & I_{zz} \end{bmatrix} \begin{Bmatrix} p\\q\\r\end{Bmatrix}\right) \\
    \begin{Bmatrix} \dot{p} \\ \dot{q} \\ \dot{r} \end{Bmatrix} = \begin{bmatrix} \frac{1}{I_{xx}} & 0 & 0 \\ 0 & \frac{1}{I_{yy}} & 0 \\ 0 & 0 & \frac{1}{I_{zz}} \end{bmatrix} \left(\begin{Bmatrix} L\\M\\N\end{Bmatrix}-\begin{bmatrix} 0 & -r & q \\ r & 0 & -p \\ -q & p & 0 \end{bmatrix} \begin{Bmatrix} I_{x}p\\I_{y}q\\I_{z}r\end{Bmatrix}\right)\\
    \begin{Bmatrix} \dot{p} \\ \dot{q} \\ \dot{r} \end{Bmatrix} = \begin{bmatrix} \frac{1}{I_{xx}} & 0 & 0 \\ 0 & \frac{1}{I_{yy}} & 0 \\ 0 & 0 & \frac{1}{I_{zz}} \end{bmatrix} \left( \begin{Bmatrix} L\\M\\N\end{Bmatrix}-\begin{Bmatrix} -I_{y}qr +I_{z}qr\\I_{x}pr -I_{z}pr\\-I_{x}pq +I_{y}pq\end{Bmatrix}\right) \\
    \begin{Bmatrix} \dot{p} \\ \dot{q} \\ \dot{r} \end{Bmatrix} = \begin{Bmatrix} \frac{L + qr(I_y-I_z)}{I_x}\\ \frac{M + pr(I_z-I_x)}{I_y} \\ \frac{N + pq(I_x-I_y)}{I_z} \end{Bmatrix}
  \end{gathered}
\end{equation}

From the simplified equation seen in Eqn. \ref{e:pqrdot_diagonal}, a fully actuated CubeSat can be controlled independent of the size of the CubeSat in finite time. This is not the case when the CubeSat becomes underactuated. To show this effect, it is assumed that only the x and y axes have actuators, so the external moment about the z-axis, N, is set to zero. The rotational dynamics of the CubeSat simplify to the equation shown below.

\begin{equation} \label{e:underactuated}
    \begin{Bmatrix} \dot{p} \\ \dot{q} \\ \dot{r} \end{Bmatrix} = \begin{Bmatrix} \frac{L + qr(I_y-I_z)}{I_x}\\ \frac{M + pr(I_z-I_x)}{I_y} \\ \frac{pq(I_x-I_y)}{I_z} \end{Bmatrix}
\end{equation}

\noindent Notice that if the mass moment of inertia along the x and y axes are equal to one another the yaw rate derivative is zero and thus the yaw rate is constant implying an uncontrollable axis. That is, the angular acceleration, $\dot{r}$, becomes zero, and the angular velocity, r, becomes a constant. If r is a constant, the yaw rate of the CubeSat never converges to zero, and the CubeSat becomes uncontrollable according to Eqn. \ref{e:underactuated}. Thus, a 1U CubeSat will always be uncontrollable if it is underactuated and has a diagonal inertia matrix. This is because a 1U uniform CubeSat would have all axes of inertia equal to each other. However, a 2U CubeSat that is underactuated can be controllable depending on the axis that is uncontrollable. This is analogous to "rotating" the CubeSat and keeping the body frame fixed. That is, if the 2U CubeSat is upright where the x-axis points out from the large face, $I_x$ is equal $I_y$ according to Eqn. \ref{e:inertia}, but due to the cubical design of a CubeSat, $I_x$ does not equal $I_y$ if the 2U CubeSat is sideways where the x-axis points out from the small face. When simulating the underactuated 2U CubeSat in Section \ref{Section:Sim}, an upright and sideways orientation of the CubeSat will be used. A 2U CubeSat in both the upright and sideways orientation are shown below. Notice that the body frame is the same for both 2U CubeSats and thus the inertia changes. 

\begin{figure}[H]
\centering
\includegraphics[width=0.8\textwidth]{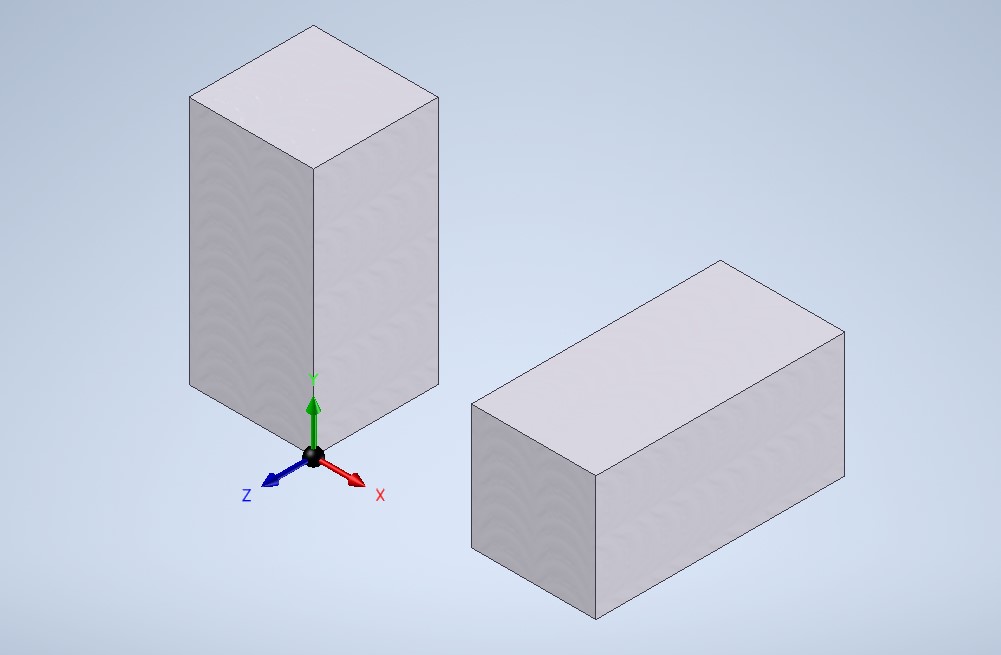}
\caption{Upright and Sideways Orientation of a 2U CubeSat}
\label{fig:2UCubeSat}
\end{figure}

\section{Control Law} \label{Section:Controllers}
%Derive P control
%Derive FL
%L=0, Iy=Iz, pdot=0, p=constant, uncontrollable
%1U - never, 2U - controllable upright but not on side, 6U works if 2-stage

In the previous section, it was shown through the dynamics that a CubeSat is fully controllable when fully actuated but is only controllable in certain configurations when underactuated. By using the dynamics discussed, the control laws to be used on the CubeSats in simulation can be derived. Three control laws will be simulated in this paper. The first control law is a proportional controller that is commonly used on CubeSats. The second control law uses feedback linearization\cite{NonlinearControl}. The third control law is a feedback linearized controller with two stages. The two stage control scheme is derived to be used for an underactuated CubeSat. 

\subsection{Proportional Control Law}
The first control law derived here is a traditional proportional controller where the error between the angular velocity vector and the desired angular velocity vector is calculated and fed back to the controller. The controller then applies proportional control to the angular velocity error to calculate the total commanded (comm) moment vector. The total moment vector is then fed into the actuators along each axis of the CubeSat. The proportional control law can be seen below where ${\bf K}_p$ is the proportional gain matrix and $K_{pi}$ is the proportional gain for the angular velocity components.

\begin{equation} \label{e:P_Control} 
  \begin{gathered}
    \vec{M}_{comm} = {\bf K}_p (\vec{\omega} - \vec{\omega}_C)\\
    {\bf K}_p = \begin{bmatrix} K_{pp} & 0 & 0 \\ 0 & K_{pq} & 0 \\ 0 & 0 & K_{pr} \end{bmatrix}
  \end{gathered}
\end{equation}

%It should be noted that the proportional controller is only viable when the CubeSat is fully actuated. From Eqn. \ref{e:pqrdot_diagonal}, the components of the angular acceleration vector will never be stuck at zero because the total moment vector is fed back into the equation. Because of this, the components angular velocity vector will never be constant. However when the CubeSat becomes underactuated as seen in Eqn. \ref{e:underactuated}, the z-axis component of the angular acceleration vector becomes permanently equal to zero if the mass moments of inertia about the x and y axes are the same, so the z component of the angular velocity vector becomes a constant. When the z component of the angular velocity vector is a constant, the CubeSat is uncontrollable.

\subsection{Feedback Linearization Control Law}
The second control law derived here is a feedback linearized controller \cite{NonlinearControl}.
%Feedback linearization is a nonlinear control technique used for systems with nonlinear dynamics. Feedback linearization takes a nonlinear control system and transforms it into an equivalent linear control system. The transformation includes a change of variables and a suitable control input  
To derive the feedback linearized controller, Eqn. \ref{e:pqrdot} needs to be linearized. First, the angular acceleration vector is set equal to a pseudo-control law, $\vec{\gamma}$.

\begin{equation}\label{e:FeedbackLinear}
  \dot{\vec{\omega}} = \vec{\gamma}
\end{equation}

\noindent Next, the pseudo-control law is substituted into Eqn. \ref{e:pqrdot} and solved for the total moment vector.

\begin{equation}\label{e:pqrdot_FL}
  \vec{M}_{comm} = {\bf I}_C \vec{\gamma} + {\bf S}(\vec{\omega}) {\bf I}_C \vec{\omega}
\end{equation}

\noindent Lastly, in order to find the total moment vector, the pseudo-control law needs to be found. The pseudo-control law applies proportional control explained in the previous section to Eqn. \ref{e:pqrdot_FL}.

\begin{equation}\label{e:pseudocontrol}
  \vec{\gamma} = {\bf K}_p (\vec{\omega} - \vec{\omega}_C)
\end{equation}

Equations \ref{e:pqrdot_FL} and \ref{e:pseudocontrol} are used together to compute the desired moments to be applied to the CubeSat. The goal of the feedback linearization control law is to cancel the nonlinear dynamics of the controller and should have better time response than the proportional control law. This feedback linearization control law is also used to derive the two stage control law in the next section. 

%It should be noted that like the proportional controller, the feedback linearized controller is only viable when the CubeSat is fully actuated. From Eqn. \ref{e:pqrdot_diagonal}, the components of the angular acceleration vector will never be stuck at zero because the total moment vector is fed back into the equation. Because of this, the components angular velocity vector will never be constant. However when the CubeSat becomes underactuated as seen in Eqn. \ref{e:underactuated}, the z-axis component of the angular acceleration vector becomes permanently equal to zero if the mass moments of inertia about the x and y axes are the same, so the z component of the angular velocity vector becomes a constant. When the z component of the angular velocity vector is a constant, the CubeSat is uncontrollable.

\subsection{Two Stage Feedback Linearized Control Law}
As shown in Equation \ref{e:underactuated}, the proportional and feedback linearized controller will not work when the CubeSat is underactuated. Because of this, another controller needs to be derived. The third control law that will be derived is a feedback linearized controller with a two stage control scheme. This control law is in line with other piece wise continuous control schemes seen in the literature \cite{Two_Control_Torque}. The two stage control scheme presented here will utilize momentum transfer between axes of the CubeSat allowing for control when the CubeSat is underactuated. Again, it is assumed that in the underactuated case, the CubeSat does not have control of the yaw axis. Since the CubeSat has actuators along the roll and pitch axes, the first stage of the two stage control scheme keeps the roll rate constant and transfers the momentum from the pitch axis to the yaw axis. Taking Eqn. \ref{e:underactuated}, the x component of the angular acceleration vector, $\dot{p}$, will be solved for first. Let the command for the x component of the angular velocity vector be equal to a constant $p_C$, and set $\dot{p}$ equal to a pseudo-control law.

\begin{equation}\label{e:FeedbackLinear_P}
  \dot{p} = \gamma_{p}
\end{equation}

\noindent Next, substitute the pseudo-control law for the x component of the angular velocity vector into Eqn. \ref{e:underactuated} and solve for the x component of the total moment vector, $L_{comm}$.

\begin{equation}\label{e:LMoment_FL}
  L_{comm} = I_x \gamma_{p} - qr(I_y - I_z)
\end{equation}

\noindent Lastly, in order to find the x component of the total moment vector the pseudo-control law needs to be found. The pseudo-control law applies proportional control explained in the previous section to Eqn. \ref{e:pqrdot_FL}.

\begin{equation}\label{e:pseudocontrol_Pdot}
  \gamma_{p} = K_{pp} (p - p_C)
\end{equation}

With the roll rate set constant, the command for the y component of the angular velocity vector can be found by solving for the z component of the angular acceleration vector, $\dot{r}$. Let the command for the z component of the angular velocity vector be equal to zero, and set $\dot{r}$ equal to a pseudo-control law.

\begin{equation}\label{e:FeedbackLinear_R}
  \dot{r} = \gamma_{r}
\end{equation}

\noindent Next, substitute the pseudo-control law for the z component of the angular velocity vector into Eqn. \ref{e:underactuated} and solve for the command for the y component of the angular velocity vector.

\begin{equation}\label{e:NMoment_FL}
  q_C = \frac{I_z \gamma_{r}}{p(I_x - I_y)}
\end{equation}

\noindent Lastly, in order to find the command for the y component of the angular velocity vector the pseudo-control law needs to be found. The pseudo-control law applies proportional control explained in the previous section to Eqn. \ref{e:pqrdot_FL}.

\begin{equation}\label{e:pseudocontrol_Rdot}
  \gamma_{r} = K_{pr} (r - r_C)
\end{equation}

With the command for the y component of the angular velocity vector found, the y component of the angular acceleration vector, $\dot{q}$, can be solved for. Set $\dot{q}$ equal to a pseudo-control law.

\begin{equation}\label{e:FeedbackLinear_Q}
  \dot{q} = \gamma_{q}
\end{equation}

\noindent Next, substitute the pseudo-control law for the y component of the angular velocity vector into Eqn. \ref{e:underactuated} and solve for the the y component of the total moment vector, $M_{comm}$.

\begin{equation}\label{e:MMoment_FL}
  M_{comm} = I_y \gamma_{q} - pr(I_x - I_z)
\end{equation}

\noindent Lastly, in order to find the y component of the total moment vector the pseudo-control law needs to be found. The pseudo-control law applies proportional control explained in the previous section to Eqn. \ref{e:pqrdot_FL}.

\begin{equation}\label{e:pseudocontrol_Qdot}
  \gamma_{q} = K_{pq} (q - q_C)
\end{equation}

\noindent In summary, here are the steps of the two stage control law.
\ \\ 
\begin{enumerate}[itemsep=-1pt]
\item Stage 1 
\begin{enumerate}[itemsep=-1pt]
\item Choose a suitable roll rate command $|p_C|>0$
\item Use Equation \ref{e:pseudocontrol_Pdot} to compute the pseudo control $\gamma_p$ for the roll axis.
\item Use Equation \ref{e:LMoment_FL} to compute the roll moment $L_{comm}$ to drive the roll rate $p$ to the commanded roll rate $p_C$.
\item Use Equation \ref{e:pseudocontrol_Rdot} to compute the pseudo control $\gamma_r$ where the commanded yaw rate is equal to zero $r_C=0$.
\item Use Equation \ref{e:NMoment_FL} to compute the desired pitch rate command $q_C$. 
\item Use Equation \ref{e:pseudocontrol_Qdot} to compute the pseudo control for the  pitch axis $\gamma_q$.
\item Use Equation \ref{e:MMoment_FL} to compute the desired pitch moment $M_{comm}$ to drive the pitch rate $q$ to the pitch rate command $q_C$.
\item Wait until the yaw rate $r$ is close to zero $|r|<\epsilon$ where $\epsilon$ is another user defined parameter. Once this condition is met, the controller switches to Stage 2.
\end{enumerate}
\item Stage 2
\begin{enumerate}[itemsep=-1pt]
\item Set the roll rate command and pitch rate commands to zero $p_C = 0,q_C=0$
\item Use equations \ref{e:pseudocontrol_Pdot} and \ref{e:pseudocontrol_Qdot} to compute the pseudo controls for roll $\gamma_p$ and pitch $\gamma_q$.
\item Use equations \ref{e:LMoment_FL} and \ref{e:MMoment_FL} to compute the desired roll $L_{comm}$ and pitch $M_{comm}$ moments to drive the roll rate $p$ and pitch rate $q$ to zero.
\item Wait until roll $p$ and pitch $q$ rates are close to zero $|p|<\epsilon,|q|<\epsilon$. Switch back to Stage 1 if residual yaw rate $r$ is still present $|r|>\epsilon$
\end{enumerate}
\end{enumerate}

%After driving the yaw rate to zero, the command for the x and y components of the angular velocity vector are set to zero, and the x and y components of the total moment vector are solved for with the new commands. The roll and pitch rates are driven to zero using the actuators along those axes. This two stage control scheme allows for control of the CubeSat when it is underactuated. However, this control law does not work for a CubeSat where the mass moments of inertia along the x and y axes are the same as previously explained in Section \ref{Section:6DOF}.

\section{Simulation Results} \label{Section:Sim}
%P Control and FL works for 1U,2U upright/side,6U if rank=3
%P Control and FL 1U,2U upright/side,6U if rank=2
%2-stage 1U,2U upright/side,6U if rank=2
%Make table for all control laws

%With the multiple control laws explained in detail, 
Simulation results of the 1U, 2U, and 6U CubeSat using the control laws from the previous section were performed. The simulation software used to simulate the CubeSat is the Facility for Aerospace Systems and Technologies Configurable Autopilot and Simulation Software Tool (FASTCASST). FASTCASST is an open-source software suite written in the C++ programming language and has multiple modes to allow for extensive testing of both software and hardware. The full software suite is stored on GitHub as a repository \cite{FASTCASST_Repo}. The modes of FASTCASST include simulation only (SIMONLY) mode, software-in-the-loop (SIL) simulation mode, hardware-in-the-loop (HIL) mode, and an autopilot mode (AUTO) for real hardware testing. The SIMONLY mode was used to simulate this CubeSat but could be used in AUTO mode if hardware testing is done in the future.  

The simulation environment contains vehicle dynamics and controls as well as external forces from gravity of the Earth. The translational dynamics are integrated using point mass physics and a point mass model for the gravitational field of the Earth. The rotational dynamics were explained in a previous section however the tranlational dynamics were not included since they do not effect the performance of the controller. Note that the attitude dynamics are done using quaternions to remove any issues with gimbal lock.  The dynamics and state of the simulated vehicle are recorded and saved upon completion of the simulation. Further explanation of the software architecture and simulation dynamics can be seen in the previous works \cite{FASTKit_Aviation,MJ_Thesis}.

The initial conditions of the CubeSat are saved in a text file and are read into FASTCASST at the start of the simulation. The initial conditions of the CubeSat are shown in the table below. The CubeSat is simulated in a low Earth orbit (LEO) of 500 kilometers. The inertial frame is about the center of the Earth, hence the large initial x-coordinate. The orbit is a circular orbit about the equator of the Earth. It should be noted that the simulation is only run for 10 seconds thus the CubeSat does not move very much in its orbit. As such the translational states are not shown in the simulations below. 

\begin{table}[H]
 \centering
 \caption{Initial Conditions of the CubeSat}
\begin{tabular}{ |c|c|  }
 \hline
 \multicolumn{2}{|c|}{\textbf{Initial Conditions for Six DOF Model with Quaternions}} \\
 \hline
 6871393.0 & Initial X Position (m) \\
 \hline
 0.0 & Initial Y Position (m) \\
 \hline
 0.0 & Initial Z Position (m) \\
 \hline
 1.0 & Quaternion 0 \\
 \hline
 0.0 & Quaternion 1 \\
 \hline
 0.0 & Quaternion 2 \\
 \hline
 0.0 & Quaternion 3 \\
 \hline
 0.0 & Initial Xbody Velocity (m/s) \\
 \hline
 7616.18 & Initial Ybody Velocity (m/s) \\
 \hline
 0.0 & Initial Zbody Velocity (m/s) \\
 \hline
 0.2 & Initial Roll Rate (rad/s) \\
 \hline
 0.2 & Initial Pitch Rate (rad/s) \\
 \hline
 0.2 & Initial Yaw Rate (rad/s) \\
 \hline
\end{tabular}
\label{t:IC_CubeSat}
\end{table}

\noindent The inertial refernce frame of the CubeSat is Earth-Centered Inertial (ECI) with the x-axis points towards the point on the surface of the Earth where the prime meridian and equator intersect. The initial x position also accounts for the radius of the Earth, 6371393 meters. This means the CubeSat is initially 500 km above the surface of the Earth as stated previously. The z-axis of the ECI frame points out of the North pole and the y-axis completes the triad. The initial angular velocities were chosen from a typical deployment rate of 10 deg/s upon ejection from a launch vehicle. The y component of velocity is calculated from the orbit of the CubeSat to be placed in a perfectly circular orbit around the equator. The only parameters that change between the different size CubeSats are the mass moments of inertia and the total mass. The mass moments of inertia and the total mass of each CubeSat is shown in the table below. 

\begin{table}[H]
 \centering
 \caption{CubeSat Mass Moment of Inertia and Total Mass}
\begin{tabular}{ |c|c|c|c|c|  }
 \hline
 \textbf{CubeSat Size} & Mass (kg) & \textbf{$I_{x} (kg \cdot m^{2})$} & \textbf{$I_{y} (kg \cdot m^{2})$} & \textbf{$I_{z} (kg \cdot m^{2})$} \\
 \hline
 \textbf{1U} & 2 & 0.0033 & 0.0033 & 0.0033 \\
 \hline
 \textbf{2U, Upright} & 4 & 0.0167 & 0.0167 & 0.0067 \\
 \hline
 \textbf{2U, Sideways} & 4 & 0.0067 & 0.0167 & 0.0167 \\
 \hline
 \textbf{6U} & 12 & 0.13 & 0.10 & 0.05 \\
 \hline
\end{tabular}
\label{t:Masses_Intertias}
\end{table}

The actuators used on these CubeSats are modeled after a thruster which will provide three-axis control. However, the thruster will be limited to two-axis control when specified. The CubeSat thruster modeled is the VACCO Integrated Propulsion System (IPS) \cite{gimballed_thruster}. The IPS is a complete propulsion module with a plug-and-play capability. The IPS has four individual thrusters that are pointed inward 7 degrees off center. With each thruster pointed off center, three-axis control is possible but is not used when the satellite is underactuated. Each thruster provides 1 newton of thrust for a total axial thrust of 4 newtons and a total impulse of 12,000 newton-seconds \cite{gimballed_thruster}. An image of the thruster is shown below.

\begin{figure}[H]
\centering
\includegraphics[width=0.6\textwidth]{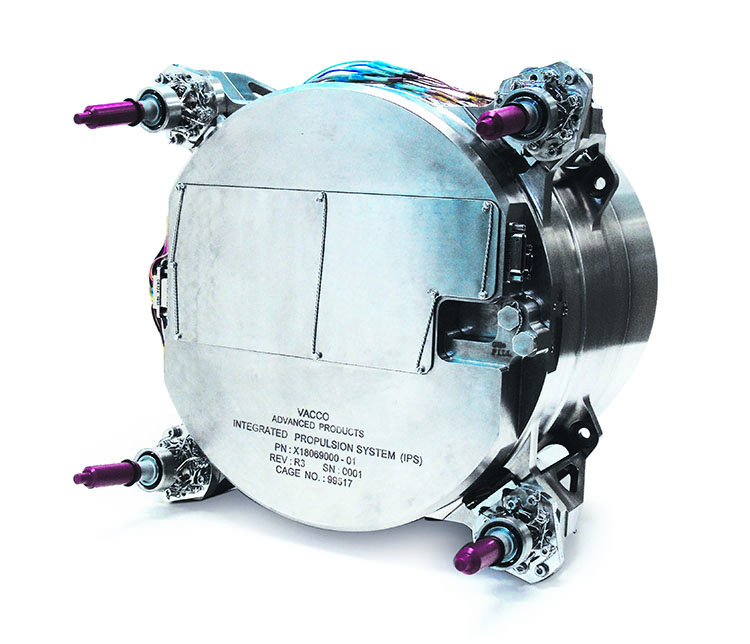}
\caption{VACCO Integrated Propulsion System\cite{gimballed_thruster}}
\label{fig:VaccoThruster}
\end{figure}

Table \ref{t:ControlScheme} shown below gives a quick summary of which control laws succeeded in driving all angular rates to zero for the different CubeSat configurations. Boxes marked with green check marks show that the control law was successful in nulling all angular rate for the CubeSat configuration while boxes marked with red Xs show that the control la was not successful. For the fully actuated CubeSat where the rank of the moment vector $\vec{M}$ is equal to three ($r(\vec{M})=3$), the proportional and feedback linearized control laws were successful in driving all angular rates to zero. When using the Two-Stage control law however, the method was only successful for the 2U Sideways Case and the 6U case. This is because of the presence of momentum transfer between axes due to $I_x$ and $I_y$ being different. For the 2U Upright case and the 1U case, these axes of inertia are equal and thus controllability becomes rank deficient and the system does not succeed in nulling all angular rates. For the scenario where the rank of the moment vector $\vec{M}$ is equal to two ($r(\vec{M})=2$), all control laws failed to drive all angular rates to zero except for the Two stage control law for the 2U Sideways and 6U case. Again the two stage control law is not effective when $I_x=I_y$ (1U, 2U Upright) but it is successful with the 2U Sideways case and the 6U case. 

%were simulated. When the CubeSat was underactuated, the proportional, feedback linearized, and two-stage control laws were used. The two-stage control law was design for an underactuated CubeSat which is why it was not simulated for a fully actuated CubeSat. The results shown in the table are shown and discussed in the following Subsections.

\begin{table}[H]
 \centering
 \caption{Control Law and CubeSat Configurations}
\begin{tabular}{ |c|c|c|c|  }
 \hline
 \multicolumn{4}{|c|}{\textbf{Fully Actuated, $r(\vec{M})=3$}} \\
 \hline
 \textbf{CubeSat Size} & \textbf{Proportional Control} & \textbf{Feedback Linearized Control} & \textbf{Two-Stage Control} \\
 \hline
 \textbf{1U} & \cellcolor{green} \checkmark & \cellcolor{green} \checkmark & \cellcolor{red} X \\
 \hline
 \textbf{2U, Upright} & \cellcolor{green} \checkmark & \cellcolor{green} \checkmark & \cellcolor{red} X \\
 \hline
 \textbf{2U, Sideways} & \cellcolor{green} \checkmark & \cellcolor{green} \checkmark & \cellcolor{green} \checkmark \\
 \hline
 \textbf{6U} & \cellcolor{green} \checkmark & \cellcolor{green} \checkmark & \cellcolor{green} \checkmark \\
 \hline
 \multicolumn{4}{|c|}{\textbf{Underactuated Actuated, $r(\vec{M})=2$}} \\
 \hline
 \textbf{CubeSat Size} & \textbf{Proportional Control} & \textbf{Feedback Linearized Control} & \textbf{Two-Stage Control} \\
 \hline
 \textbf{1U} & \cellcolor{red} X & \cellcolor{red} X & \cellcolor{red} X \\
 \hline
 \textbf{2U, Upright} & \cellcolor{red} X &  \cellcolor{red} X & \cellcolor{red} X \\
 \hline
 \textbf{2U, Sideways} & \cellcolor{red} X &  \cellcolor{red} X & \cellcolor{green} \checkmark \\
 \hline
 \textbf{6U} & \cellcolor{red} X &  \cellcolor{red} X & \cellcolor{green} \checkmark \\
 \hline
\end{tabular}
\label{t:ControlScheme}
\end{table}

\begin{figure}[H]
  \begin{center}
    \begin{tabular}{cc}
      \includegraphics[height=100mm,width=120mm]{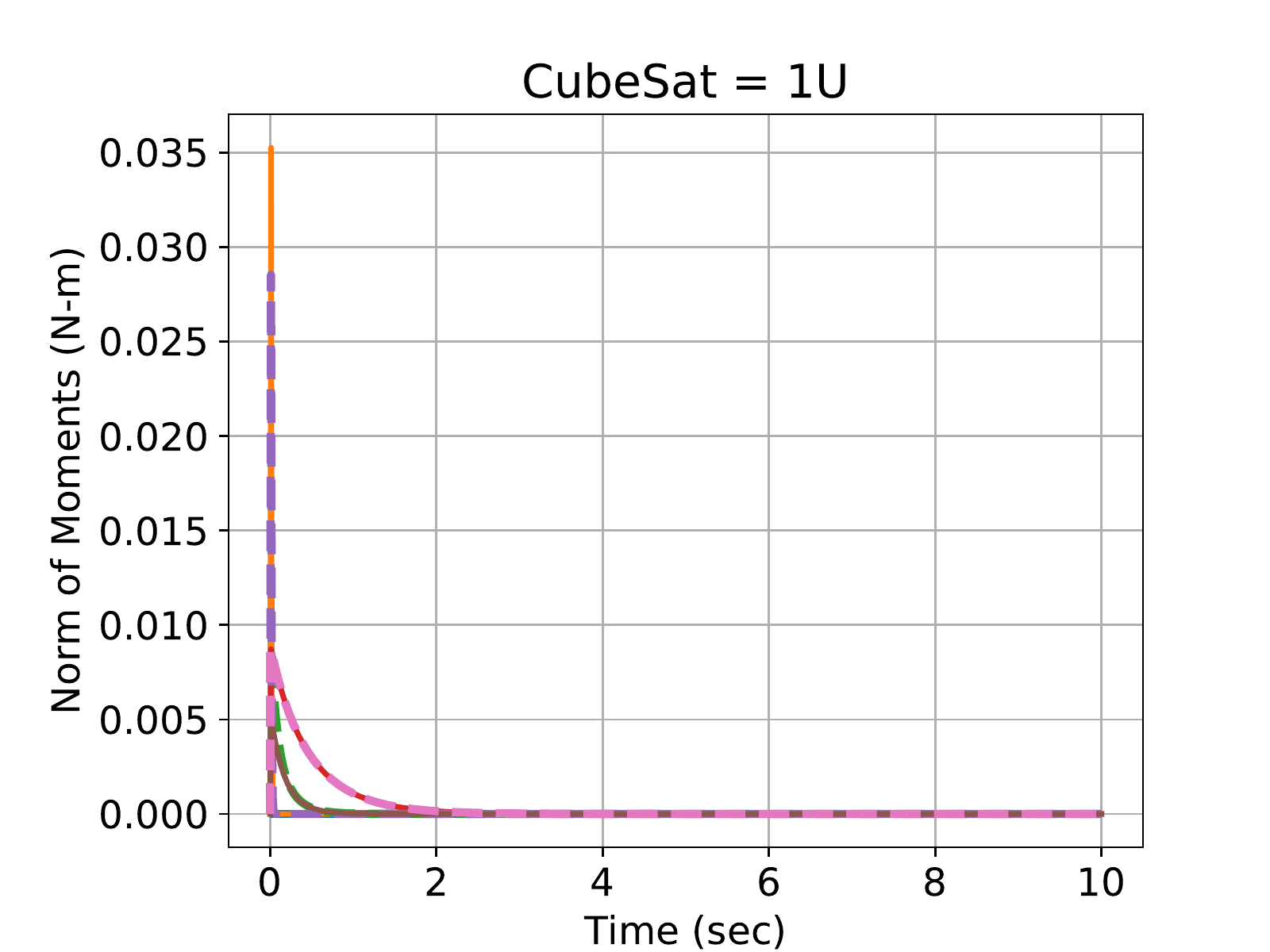}
      &
      \includegraphics[height=35mm,width=35mm]{Figures/CubeSAT_Results/Legend.png} \\
    \end{tabular}
  \end{center}
\caption{\bf 1U CubeSAT Norm of External Moment (N-m) vs Time(sec)}
  \label{f:1UM3Moment}
\end{figure}

Simulation results are shown for a few CubeSat configurations. First, the 1U CubeSat is shown in the Figure below. The first line in the legend is the dashed blue line which represents the No Control case. In this scenario the CubeSAT tumbles at it's initial velocity of 0.2 rad/s along each axis (norm = 0.35 rad/s). Since no moments are placed on the spacecraft, the norm stays constant due to conservation of angular momentum. The orange line and green dashed lines represent the Proportional-Integral-Derivative (PID) control law as well as the Feedback Linearization (FL) control laws for the fully actuated case (($r(\vec{M})=3$). In this case after about 1.5 seconds, the norm of the angular velocity is zero. The red solid line and the pink dashed lines represent the Two Stage controller for both fully actuated and underactuated but the angular velocity of the system actually increases. This is because the first stage of the control law requires the roll rate to reach a steady state of 0.5 rad/s. Since there is no momentum transfer between axes the roll rate and pitch rates reach steady state and the yaw rate never changes. The purple dashed line and solid brown line represents the PID and FL cases respectively when the actuator is not full rank ($r(\vec{M})=2$). In this case the controllers attempt to drive the angular velocity to zero but the yaw rate remains unchanged at 0.2 rad/s and thus the norm remains at 0.2 rad/s in steady state. 
\begin{figure}[H]
  \begin{center}
    \begin{tabular}{cc}
      \includegraphics[height=100mm,width=120mm]{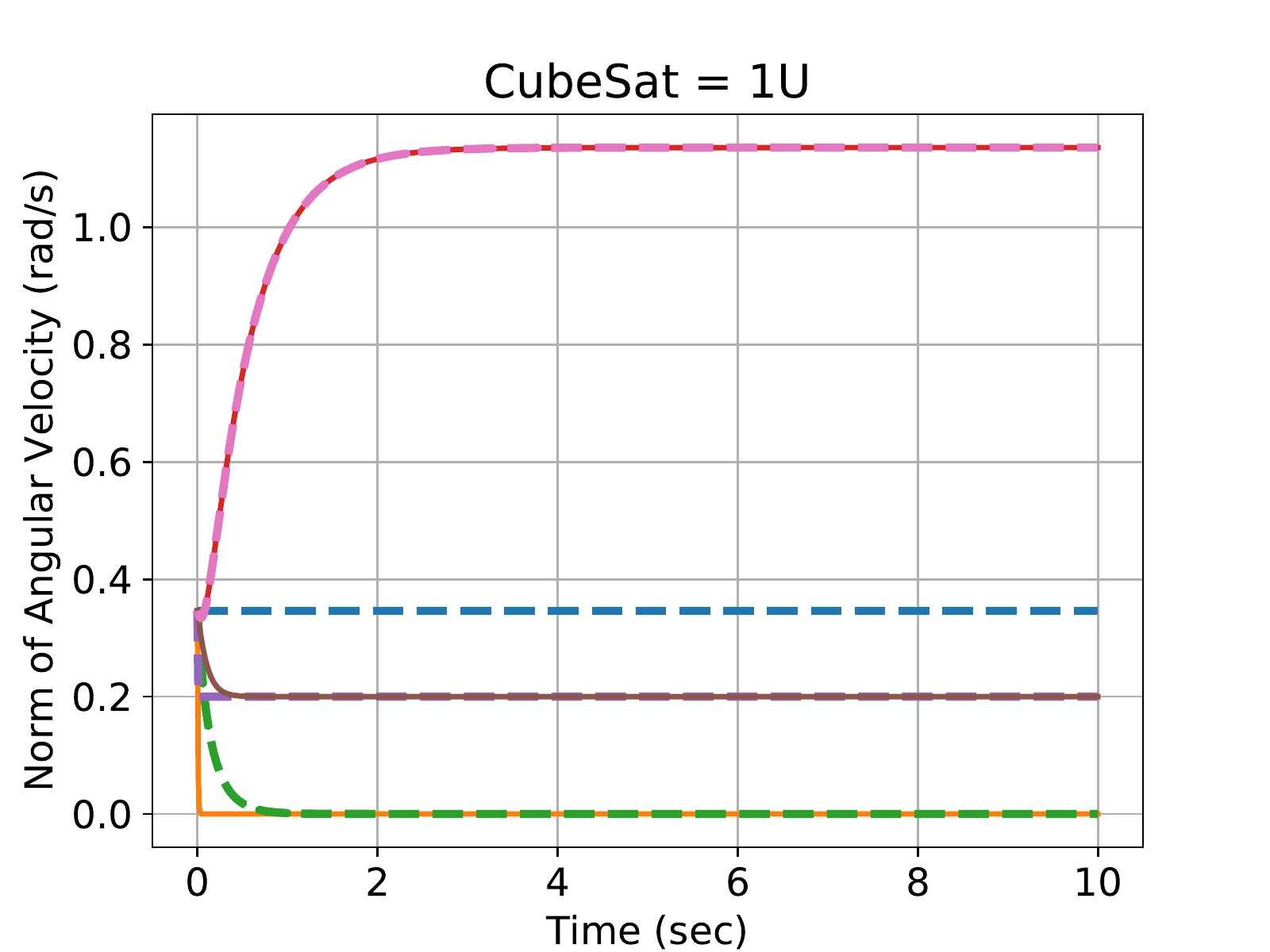}
      &
      \includegraphics[height=35mm,width=35mm]{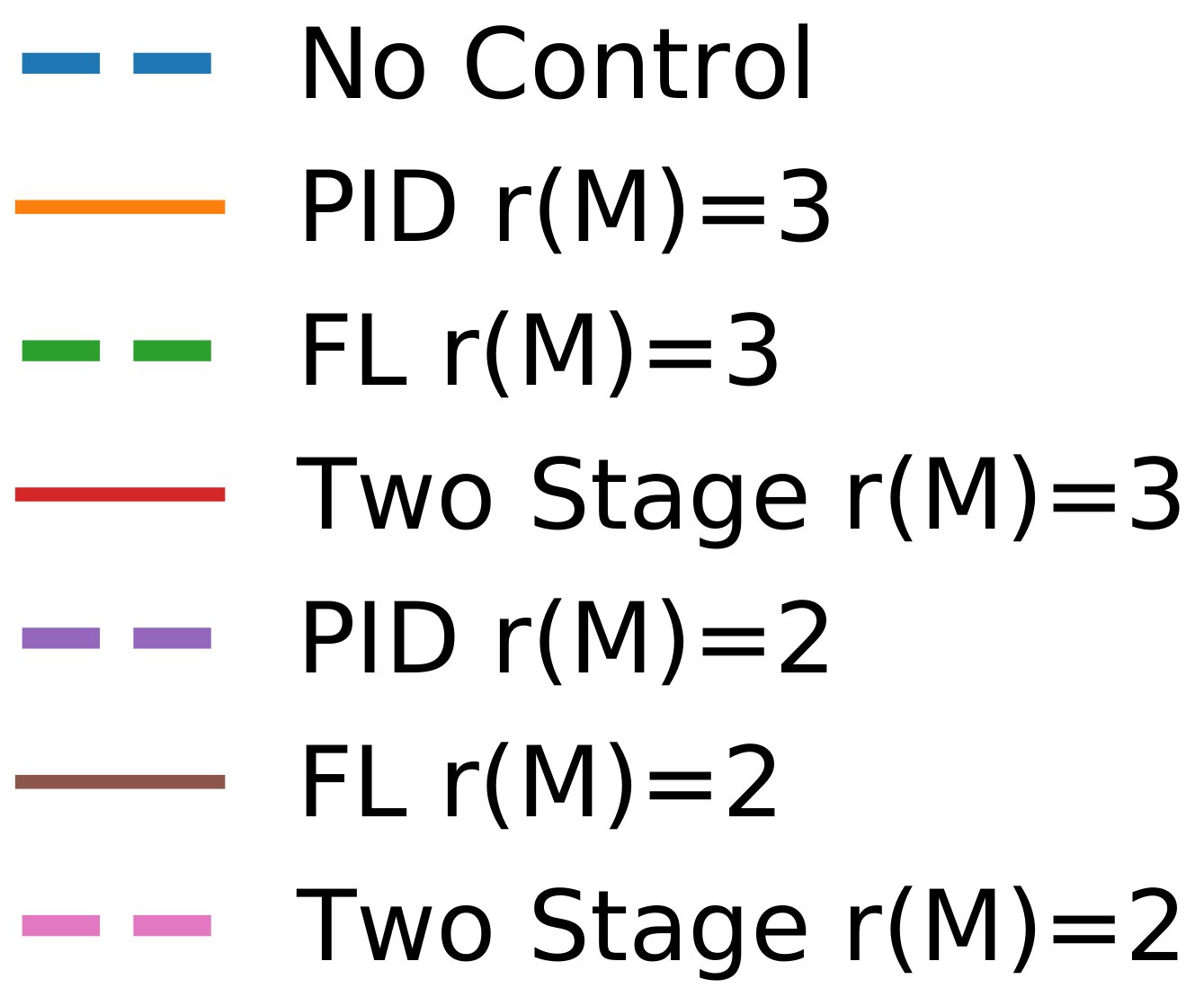} \\
    \end{tabular}
  \end{center}
\caption{\bf 1U CubeSAT Norm of Angular Velocity (rad/s) vs Time(sec)}
  \label{f:1UM3}
\end{figure}
Notice also, that the two stage control time response is the same whether the actuators are full rank or not. This is because the controller does not take advantage of the 3rd control axis in either case. Figure \ref{f:1UM3Moment} shows the total moment applied to the 1U CubeSAT for all 3 control types and actuator ranks. In all 3 scenarios the moments eventually arrive at zero due to the fact that each control law reaches steady state within about two seconds. 
Similar results can be produced for the 2U Upright case since $I_x=I_y$. The control law time responses change when $I_x$ does not equal $I_y$ however. Such is the case for the 2U Sideways and 6U cases. The results for the 2U Sideways case are shown below. Once again the dashed blue line represents no control. In this case the norm of the angular velocity remains constant at 0.35 rad/s. The orange solid line and green dashed lines represent the PID and FL control laws for three axis control. 
\begin{figure}[H]
  \begin{center}
    \begin{tabular}{cc}
      \includegraphics[height=100mm,width=120mm]{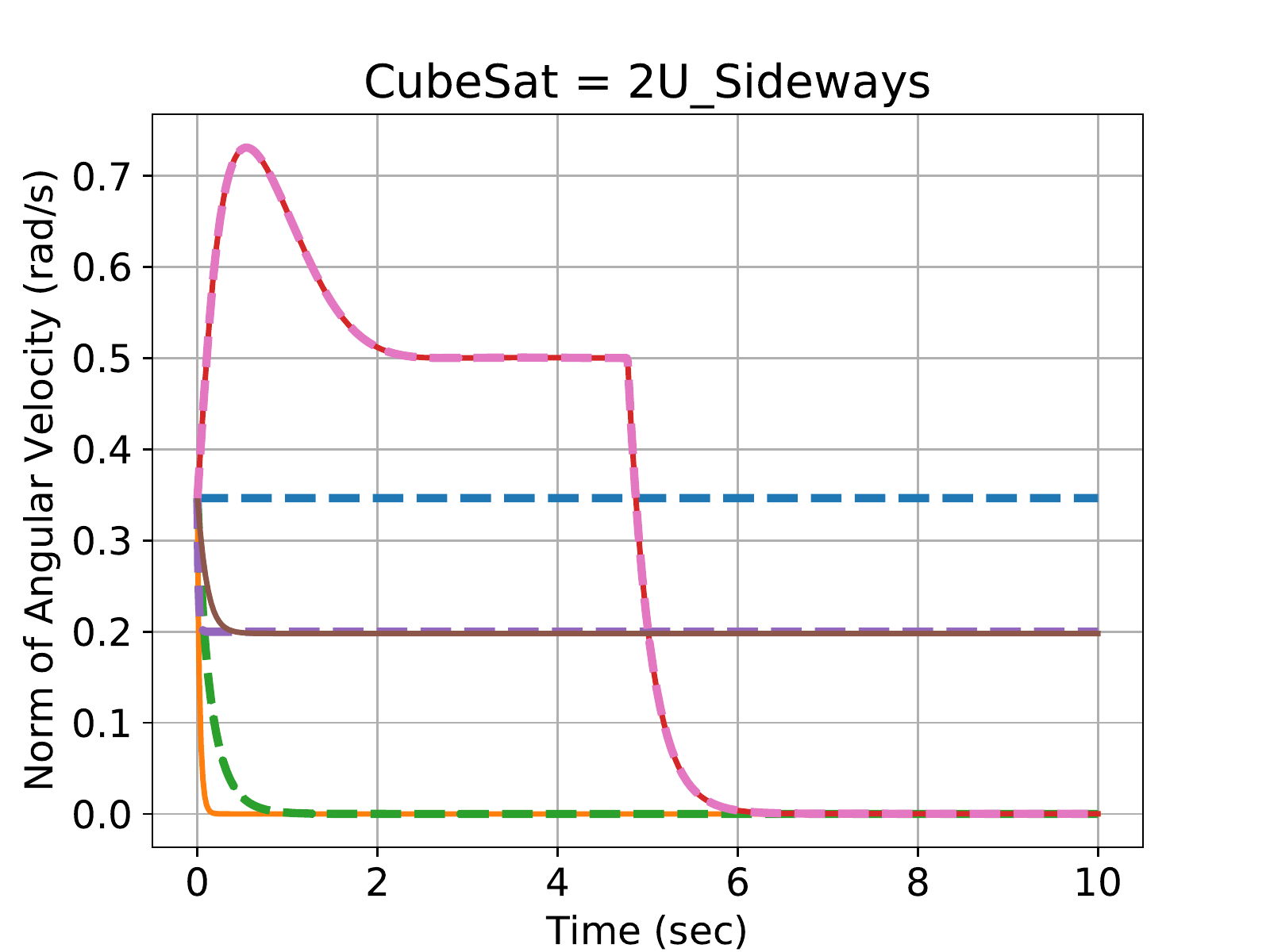}
      &
      \includegraphics[height=35mm,width=35mm]{Figures/CubeSAT_Results/Legend.png} \\
    \end{tabular}
  \end{center}
\caption{\bf 2U Sideways CubeSAT Norm of Angular Velocity (rad/s) vs Time(sec)}
  \label{f:2UM3}
\end{figure}
The next 3 lines represent the control law time response when the yaw moment is zero. In this case the PID and FL control laws cannot drive the yaw axis to zero. This is seen in the purple dashed line and solid brown line. The steady state angular velocity value is 0.2 rad/s again due to the fact that the yaw rate remains unchanged at 0.2 rad/s. However, when the two stage control law is used (solid red and dashed pink) the control law is successful in nulling the rates to zero. In this case the pitch rate is commanded to a constant 0.5 rad/s. It takes around 5 seconds for the pitch rate to reach 0.5 rad/s and the yaw rate to reach close enough to zero for the control law to switch from stage 1 to stage 2. The discontinous jump at 5 seconds is a consequence of the piece-wise continous control law derived earlier. 
\begin{figure}[H]
  \begin{center}
    \begin{tabular}{cc}
      \includegraphics[height=100mm,width=120mm]{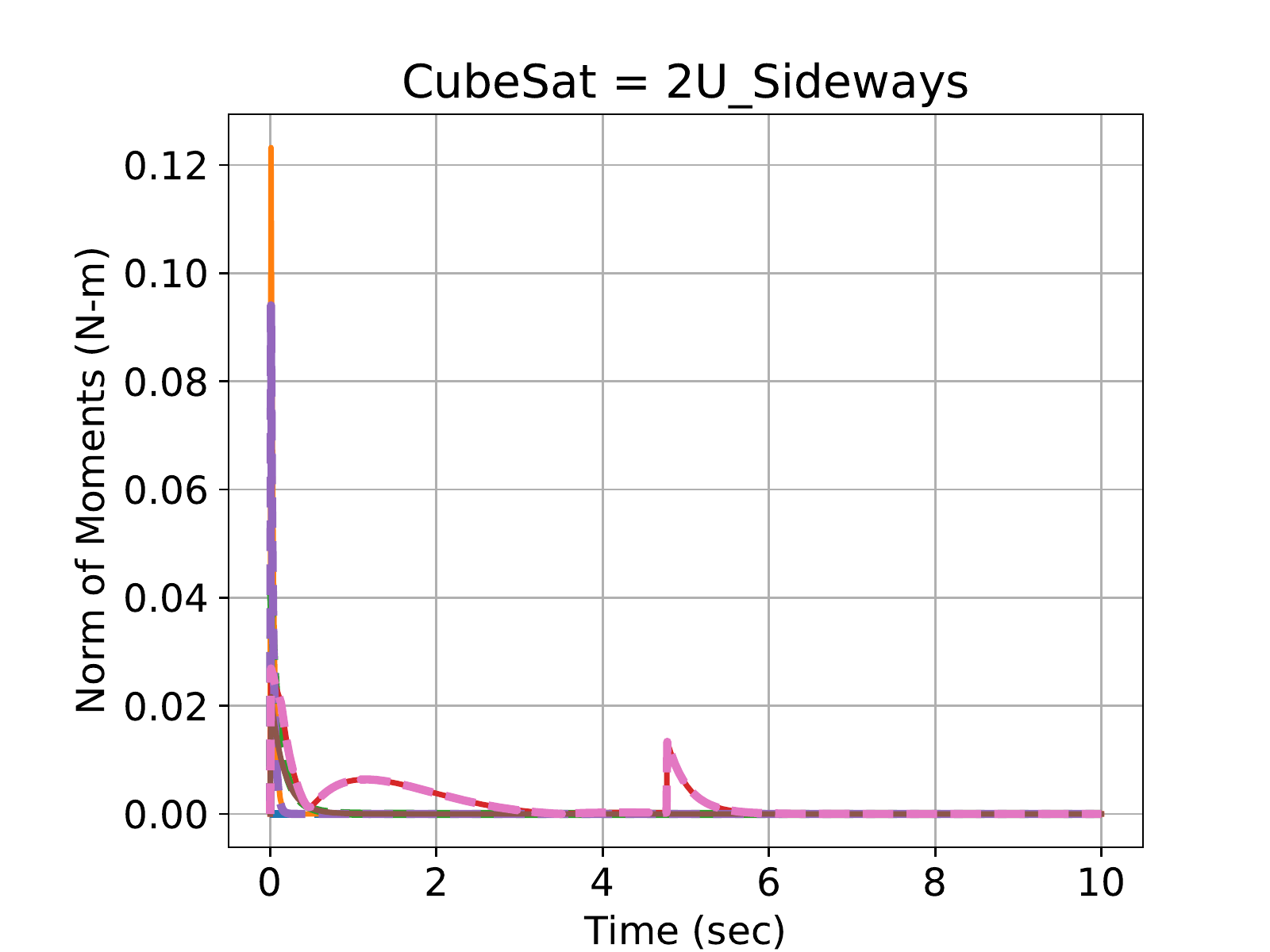}
      &
      \includegraphics[height=35mm,width=35mm]{Figures/CubeSAT_Results/Legend.png} \\
    \end{tabular}
  \end{center}
\caption{\bf 2U Sideways CubeSAT Norm of External Moment (N-m) vs Time(sec)}
  \label{f:2UM3Moment}
\end{figure}
The norm of all moments are then shown in Figure \ref{f:2UM3Moment}. The figure is similar to the 1U case except at 5 seconds a discontinuous jump is seen in the two stage control law case when the control switches from stage 1 to stage 2. In steady state the angular velocity vector and moment vector reach zero thus successfully nulling the angular rates. Similar results are seen for the 6U case but are not shown for brevity. 

\section{Conclusion and Future Work}
This paper details the design of an underactuated control law for different CubeSat configurations. The dynamics and mass moments of inertia of a 1U, 2U, and 6U CubeSat were explored before deriving a proportional and a feedback linearized control law. Once the control laws were derived, a two-stage control scheme was developed for the feedback linearized control law to control an underactuated CubeSat where only 2 axes contain actuators. All control laws were tested on each CubeSat configuration and the results were compared.

The simulation results presented in Section \ref{Section:Sim} show the verification and validation of the control laws. The simulation software used was the open source modeling software FASTCASST. Both the proportional and feedback linearized controllers were shown to work on different size CubeSats if the CubeSats were fully actuated. However, when the different size CubeSats become underactuated both of the controllers become unsuccessful in driving all angular rates to zero. To make the feeback linearized controller succesful for an underactuated CubeSat, the two-stage control scheme was implemented, and the control scheme was shown to work. Table \ref{t:ControlScheme} shows which control laws worked for which CubeSat and it is clear that when the moments of inertia along the x and y axes are not equal the two stage control law is successful. 

Future work for the feedback linearized control law include further development of the two-stage control scheme and further development of the simulation. With the two-stage feedback linearized control law shown to have worked with an underactuated CubeSat with a diagonal inertia matrix, the control law can be further tested on a CubeSat with a non-diagonal inertia matrix. As shown, certain CubeSat sizes are not controllable when underactuated. By using a non-diagonal inertia matrix, it could become possible to control a CubeSat regardless of the size, but further exploration of the CubeSat dynamics is needed to confirm this. As previously stated with the two-stage control scheme working on an underactuated CubeSat, the simulation can be further developed to include disturbance torques during the orbit of the CubeSat. When modeling the disturbance torques, the two-stage control law can be tested to see if it can desaturate reaction wheels increasing the capabilities of the controller.

%\section*{Appendix}
%\subfile{Sections/Appendix}

%\section*{Acknowledgments}
%\subfile{Sections/Acknowledgements}

\bibliography{smallsat}
\end{document}